# Low-loss far-infrared surface phonon polaritons in suspended SrTiO$_3$ nanomembranes

Konnor Koons[1], Hans A. Bechtel[2], Javier Taboada-Gutiérrez [3], Reza Ghanbari[1], Yueyin Wang[1], Stephanie N. Gilbert Corder[2], Alexey B. Kuzmenko[3], Ruijuan Xu[1*], Yin Liu[1*]

1. Department of Materials Science and Engineering, North Carolina State University, Raleigh, NC 27695, United States
2. Advanced Light Source Division, Lawrence Berkeley National Laboratory, Berkeley, CA 94720, United States
3. Department of Quantum Matter Physics, University of Geneva, 1211 Geneva, Switzerland

*Emails: rxu22@ncsu.edu and yliu292@ncsu.edu

**Abstract:**

Phonon polaritons (PhPs), excitations arising from the coupling of light with lattice vibrations, enable light confinement and local field enhancement, which is essential for various photonic and thermal applications. To date, PhPs with high confinement and low loss have been mainly observed in the mid-infrared regime and mostly in manually exfoliated flakes of van der Waals (vdW) materials. In this work, we demonstrate the existence of low-loss, thickness-tunable phonon polaritons in the far-infrared regime within transferable free-standing SrTiO$_3$ membranes synthesized through a scalable approach, achieving high figures of merit, which are comparable to the previous record values from the vdW materials. Leveraging atomic precision in thickness control, large dimensions, and compatibility with mature oxide electronics, functional oxide membranes present a promising large-scale two-dimensional (2D) platform alternative to vdW materials for on-chip polaritonic technologies in the far-infrared regime.

**Main text**

**Introduction**
Phonon polaritons (PhPs) are hybrid optical modes arising from the coupling of photons with optical phonons either confined in the bulk or at the surface of polar materials. Specifically, surface phonon polaritons (SPhPs) can be excited in Reststrahlen bands (RBs), the spectral regions between the transverse and longitudinal optical phonon frequencies, where the real part of the dielectric function is negative[1]. The ability of PhPs to compress the light wavelength and enhance optical fields at the subdiffractional scale has enabled various applications including optical sensing[2,3], light guiding[4–6], perfect absorption[7], superlensing[8–10], coherent thermal emission[11], and heat transfer[12–16]. Recent studies have focused extensively on the two-dimensional van der Waals (vdW) materials, such as hexagonal boron nitride (hBN)[4,17–21], molybdenum trioxide ($MoO_3$)[22–29], and vanadium pentoxide ($V_2O_5$)[30], supporting highly confined PhPs with low propagation losses. However, high-quality vdW material samples are typically fabricated by mechanical exfoliation techniques. This stochastic method offers limited control over lateral size and thickness, preventing scalability and practical device integration. While large vdW materials can be produced via scalable chemical vapor deposition (CVD), their PhP application is challenged by high defect densities and misaligned grains, which introduce phononic losses and degrade polariton propagation[27,31,32].

Another factor limiting the use of PhPs is that the available low-loss polaritonic materials, such as the aforementioned vdW compounds[30,33,34] and covalent silicon carbide (SiC)[1], primarily operate in the mid-infrared range (above 750 cm$^{-1}$). Experimental demonstrations of highly confined PhPs in the far-infrared and terahertz regime that propagate over a distance of several wavelengths are scarce, primarily focusing on terahertz PhPs in vdW materials including GeS[35,36] and $MoO_3$[37], with the figures of merit being moderate compared to their mid-infrared counterparts. This presents a significant obstacle for crucial applications of PhPs for long-wavelength optoelectronics and thermal management. Specifically, PhPs have demonstrated the ability to enhance the in-plane thermal conductivity of free-standing films or nanowires made from polar dielectrics such as SiC[13], $SiN_x$[38] and $SiO_2$[14,39], providing an intriguing approach for heat dissipation in microelectronics and integrated photonics. However, these applications necessitate low-loss far-infrared PhPs capable of being thermally excited at relatively low temperatures. The conductivity enhancement in these materials remains limited at room temperature either due to the high loss of far-infrared PhPs in $SiN_x$ and $SiO_2$ or the low population of thermally excited mid-infrared PhPs in SiC.

To overcome the limitations of the random fabrication and constrained range of operating frequencies, it is essential to explore scalable alternative materials that support low-loss, tunable PhPs in the far-infrared range (below 400 cm$^{-1}$ or equivalently with free-space wavelength longer than 25 μm). Recently, high-quality, free-standing complex oxide membranes have been created[40–53] and explored as a novel platform for infrared PhPs[54,55]. Compared to vdW materials, these membranes, synthesized through layer-by-layer thin-film epitaxy followed by chemical lift-off, are more scalable and reproducible, with millimeter-scale lateral dimensions. The synthesis process provides remarkable crystalline quality and enables atomic-scale thickness control, allowing for precise adjustments ranging from a single-unit cell thickness to several hundred nanometers[40–45]. This level of precision greatly facilitates the fine-tuning of the dispersion relation of PhPs by controlling the thickness. Our recent work has demonstrated that perovskite $SrTiO_3$ membranes exhibit highly confined SPhPs in the mid-infrared regime[54]. Despite this promise, far-infrared PhPs in the oxide membranes have not been investigated in the previous study due to significant optical losses inherent to the $SiO_2$ on Si substrate, which possesses a broad lossy phonon mode at around 450 cm$^{-1}$ hindering PhP propagation[54].

In this work, we investigate the far-infrared SPhPs in suspended $SrTiO_3$ membranes with the thickness of 30-100 nm using near-field synchrotron infrared nanospectroscopy (SINS) (Fig.1a). The suspension creates a lossless and symmetric dielectric environment for the membranes, enabling exploration of the PhP propagation behavior limited only by intrinsic losses. We analyze position-dependent SINS spectra to determine the complex-valued polariton wavevector $k_p$ (and the dispersion relations) and propagation length, which agree well with our theoretical calculations and simulations. Our findings reveal a high polaritonic quality factor $Q = \frac{\text{Re}(k_p)}{\text{Im}(k_p)}$ up to 24 and an exceptionally long lifetime up to 8 picoseconds for the far-infrared SPhPs in the 100 nm thick $SrTiO_3$ membranes. These values are comparable to the Q-factor and lifetime of mid-infrared PhPs in low-loss polaritonic materials such as hBN, SiC and $MoO_3$ and surpass that of terahertz PhPs in $MoO_3$ and GeS. Furthermore, we demonstrate the possibility to tune the polariton dispersion by varying the thickness of the membranes from 30 to 100 nm. We verify that a similarly high $Q$ factor and lifetime can also be achieved in 50 nm thick membranes, where are notably 500 times thinner than the free-space wavelength. Our work unlocks the potential of complex functional oxide membranes for advanced far-infrared and terahertz polaritonic applications.

**Theoretical calculations of SPhPs in suspended $SrTiO_3$ membranes**
Since $SrTiO_3$ has a cubic crystal structure with isotropic optical properties, for our theoretical calculations, we use the isotropic dielectric function of bulk $SrTiO_3$, $\epsilon_{STO}(\omega)$, extracted in our previous work from the far-field Fourier-transform infrared (FTIR) spectroscopy on bulk crystals[54]. Our analysis of the far-field reflectivity spectra of the membranes confirms the validity of using the dielectric function of bulk $SrTiO_3$ at least for 100 nm thick $SrTiO_3$ membrane (Supplementary Note 1 and Supplementary Fig.1)[54]. While it is challenging to extract the dielectric function from the FTIR spectra on thinner membranes (since they are dominated by the substrate), we adopt an approximation in our further numerical simulations that the dielectric function is thickness independent. $\epsilon_{STO}(\omega)$ displays three RBs corresponding to three sets of TO and LO modes (Fig.1b). The RBs cover from the far-infrared (terahertz) to the mid-infrared regime, enabling a broad range of operating frequencies of SPhPs, with RB$_1$, RB$_2$ and RB$_3$ ranging from 95-172 cm$^{-1}$, 175-475 cm$^{-1}$ and 546-798 cm$^{-1}$ respectively. These results are consistent with previously reported infrared dielectric function of SrTiO$_3$[56,57]. Our work particularly focuses on the SPhPs in RB$_2$ and RB$_3$ (highlighted in Fig.1b) that can be experimentally probed by SINS.

We calculate the dispersion relationship and quality factor of SPhPs in suspended SrTiO$_3$ membranes. The symmetric and antisymmetric modes (as determined by the parity of the E$_z$ field component) are formed due to the interaction of the SPhPs at the upper and lower membrane surface. The dispersions of these two modes in a membrane of thickness $t$ suspended in air are given by the following implicit relations[58,59],

$$\sqrt{\frac{k_{sym}^2(\omega,t)-k_0^2(\omega)}{k_{sym}^2(\omega,t)-k_0^2(\omega)\epsilon_{STO}(\omega)}} = \frac{1}{\epsilon_{STO}(\omega)} \times \tanh\left(-\frac{t}{2}\sqrt{k_{sym}^2(\omega,t)-k_0^2(\omega)\epsilon_{STO}(\omega)}\right) \quad (1)$$

$$\sqrt{\frac{k_{asym}^2(\omega,t)-k_0^2(\omega)}{k_{asym}^2(\omega,t)-k_0^2(\omega)\epsilon_{STO}(\omega)}} = \frac{1}{\epsilon_{STO}(\omega)} \times \coth\left(-\frac{t}{2}\sqrt{k_{asym}^2(\omega,t)-k_0^2(\omega)\epsilon_{STO}(\omega)}\right) \quad (2)$$

where $k_0(\omega) = \omega/c$ is the free-space wavevector, and $k_{sym}(\omega)$ and $k_{asym}(\omega)$ are the complex-valued wavevectors for the higher-energy symmetric mode and the lower-energy antisymmetric mode, respectively.

Solving equations (1) and (2) generates the complex-valued polariton wavevector $k_p$ as a function of frequency $\omega$. The calculated dispersions for the SPhPs in RB$_2$ and RB$_3$ of suspended SrTiO$_3$ are shown in Fig.1c. As the thickness is reduced, the energy splitting between the antisymmetric (solid lines) and symmetric (dashed lines) modes and the real part of the in-plane momentum $k_p$ for the antisymmetric mode increase across all frequencies in both RB$_2$ and RB$_3$. For membranes with deep-subwavelength thickness (30-100 nm), the symmetric SPhPs are non-propagating modes (also known as epsilon-near-zero (ENZ) modes[60]) exhibiting very small group velocities, whereas the antisymmetric SPhPs are propagating modes. The confinement factor $C = \frac{\text{Re}(k_p)}{k_0}$, which quantifies the compression of the wavelength in SPhPs as compared to light, increases from 10 in 100 nm membranes to over 30 in 30 nm membranes for the far-infrared antisymmetric SPhPs in RB$_2$ (Supplemental Note 2 and Supplementary Fig.2). We also calculate the $Q$ factor (Fig.1d), which characterizes the optical loss for the SPhPs propagation. We note that the expected quality factors in the bulk limit for SPhPs with little confinement are extremely high (up to $10^2$ in RB$_3$ and $10^3$ and RB$_2$) and comparable to the best SPhP materials, such as SiC[1]. The higher value in RB$_2$ indicates a longer SPhP propagation length[61]. As the thickness is reduced to submicron scales, the $Q$ factor lowers significantly while the confinement factor increases. Interestingly, when the membrane thickness is reduced below 100 nm, $\text{Re}(k_p)$ scales with $\text{Im}(k_p)$ and the $Q$ factor is nearly independent of thickness. This behavior is consistent with the results previously reported in SiC thin films[59]. Notably, in the RB$_2$ band, high Q factors peaking at 25 are obtained for far-infrared antisymmetric SPhPs, comparable to that of SPhPs in top polaritonic materials such as SiC, hBN and MoO$_3$ are observed (a detailed comparison will be provided later). In contrast, the maximum Q factor in the RB$_3$ band is only 7, consistent with the experimentally reported value of 5 in our previous study[54]. The significantly higher Q factors in RB$_2$ compared to RB$_3$ indicate a low loss of the SPhPs, consistent with a larger real-to-imaginary part ratio of $\epsilon_{STO}(\omega)$ in RB$_2$. The electric field distribution (Fig.1e) corresponding to the antisymmetric SPhPs at two characteristic frequencies in the RB$_2$ and RB$_3$ bands (obtained using finite-difference time-domain (FDTD) full-field simulations) confirms the presence of surface confined modes and indicates a longer SPhP propagation length in RB$_2$.

**Synthesis of high-quality suspended SrTiO$_3$ membranes**

For our experiments, we synthesize 30, 50, and 100 nm thick SrTiO$_3$ membranes via atomic-scale thin film epitaxy. The epitaxial heterostructures of 30-100 nm SrTiO$_3$ films with a 12 nm water-soluble Sr$_3$Al$_2$O$_6$ sacrificial layer are grown on single-crystalline (001)-oriented SrTiO$_3$ substrates via reflection high energy electron diffraction (RHEED)-assisted pulsed-laser deposition (PLD) (Methods). The synthesis of both SrTiO$_3$ and Sr$_3$Al$_2$O$_6$ layers are controlled in a layer-by-layer 2D growth mode, monitored *in situ* by RHEED (Supplementary Fig.3a). X-ray theta-2theta scans are performed on the as-grown heterostructures, revealing that the films are crystalline and free of impurity phases (Supplementary Fig.3b). By selectively dissolving Sr$_3$Al$_2$O$_6$ in deionized water, SrTiO$_3$ membranes are released from the as-grown substrates and transferred onto a SiO$_2$/Si substrate with arrays of 5 μm wide and 1 μm deep trenches, patterned via photolithography and reactive ion etching (RIE) (Methods and Fig.2a). The released membranes, which are crack-free on the millimeter lateral scale, are suspended across the gap of micro-trenches on SiO$_2$/Si substrates (Fig.2b), therefore minimizing interaction of SPhPs with the substrate. Essentially for the SINS experiment, the membranes are aligned in such a way that a sharp and straight edge is oriented perpendicular to the direction of the trenches as revealed by the SEM imaging (Fig.2c). Moreover, atomic force microscopy (AFM) imaging reveals the presence of atomic step terraces, indicating that the membrane surface is atomically smooth (Fig.2d). The released membranes remain highly crystalline, as confirmed by the X-ray $\theta$-$2\theta$ scans and rocking curve measurements showing a full width at half maximum (FWHM) value of 0.15° (Supplementary Figs.3c,d). Notably, the lattice parameter extracted from the theta-2theta scans for the SrTiO$_3$ membranes closely matches the bulk value, indicating the absence of residual strain and nonstoichiometry in the synthesized membranes (Supplementary Note 3). Atomic-resolution scanning transmission electron microscopy (STEM) imaging (Fig.2e) further confirms the cubic lattice and high crystalline quality of the synthesized membranes.

**SINS measurement of the SPhPs in the SrTiO$_3$ membranes**

We then conduct SINS to probe the infrared SPhPs in the suspended SrTiO$_3$ membranes[62]. In the SINS measurement, a broadband infrared synchrotron light source is directed onto a metal-coated AFM tip of a scattering-type scanning near-field optical microscopy (s-SNOM) setup, providing the necessary momentum for the optical excitation of the phonon polaritons. SINS enables the measurement of optical near-field spectra with nanometer resolution across a broad spectral range, which has been successfully used to probe phonon polaritons in various materials[20,33,54]. To investigate the far-infrared SPhPs, our SINS measurement employs a liquid helium-cooled HgCdTe (MCT) detector equipped with a CsI window[36], allowing us to capture near-field signals down to below 200 cm$^{-1}$.

Fig.3a shows the near-field, broadband optical image of a 100 nm thick SrTiO$_3$ membrane suspended over a trench. The suspended membrane shows a reasonable near-field signal intensity, although it is lower than that of the membrane on a SiO$_2$/Si substrate. This observation is consistent with the reduced reflectivity in the absence of the high-refractive-index substrate. A set of SINS spectra on the suspended SrTiO$_3$ are acquired at varying distances from the edge (along the green line in Fig.3b). In this study, we utilize the demodulated second-harmonic amplitude and phase signals ($s_2(\omega)$ and $\varphi_2(\omega)$), which have been normalized against the spectra of a reference gold film.

We begin with the spectrum acquired far away (20 μm) from the edge (green solid curve in Fig.3b), where the spectral response is minimally influenced by the interference of polaritons generated by the edges. It exhibits two peaks at 300 cm$^{-1}$ and 550 cm$^{-1}$ which are attributed to the excitation of the antisymmetric SPhPs in RB$_2$ and RB$_3$ respectively[54]. Notably, no symmetric (ENZ) PhP features are observed in the suspended membrane. In contrast, strong ENZ peaks are observed in SINS spectra acquired from SrTiO$_3$ supported on gold or SiO$_2$/Si substrates[54]. This can be explained by the symmetric dielectric environment of the suspended SrTiO$_3$, which allows the SINS tip to predominantly excite the antisymmetric modes (Supplementary Note 4 and Supplementary Fig.4). Further, we perform finite dipole modeling[63] to simulate the SINS spectrum based on the extracted $\epsilon_{STO}(\omega)$, assuming an infinitely large lateral size for the SrTiO$_3$ membrane. The simulated SINS spectrum reasonably matches the spectral features in RB$_2$, corroborating the absence of the ENZ mode (Supplementary Note 5 and Supplementary Fig.5). As we approach the edge, the spectra become position-dependent and reveal new peaks in both RB$_2$ and RB$_3$ as compared to the data collected far from the edge. The dashed curves in Fig.3b highlight these spatially evolving features, arising from SPhP propagation.

To analyze the dispersion and $Q$ factor of the SPhPs, we record the spectra at various points along a line perpendicular to the edge (where $x = 0$) with a step of 100 nm. Fig.3c shows the 2D plot of amplitude spectra $s_2(\omega)$ obtained from the line scan on a 100 nm thick suspended SrTiO$_3$ membrane. Multiple fringes associated with the SPhP propagation are observed in RB$_2$, while a single fringe is noted in RB$_3$. This observation is consistent with the calculated $Q$ values for SPhPs in both RB$_2$ and RB$_3$, as well as with our previous study of SPhPs in RB$_3$ in SrTiO$_3$ supported on a substrate[54]. The observed fringes result from the interference of SPhPs with the backscattered light. Depending on the illumination conditions, different SPhP paths contribute to the periodicity of the fringes[64]. The first path involves SPhPs launched by the tip, backscattered by the edge, and then scattered by the tip into the detector, introducing a periodicity corresponding to twice the polariton wavelength. The second path involves SPhPs launched by the edge and scattered by the tip into the detector, resulting in a periodicity corresponding to the polariton wavelength. To illustrate the contribution from these paths, we perform a Fourier transform of the amplitude $s_2(\omega)$ profile at different frequencies (rows of the 2D plot) in RB$_2$ (Fig.3d). The results confirm that the real-space amplitude profile contains two components with distinct spatial frequencies, corresponding to the $Re(k_p)$ and $2Re(k_p)$, which match the values calculated for the 100 nm thick sample (Fig.1c). The $2k_p$ component is attributed to the tip-launched SPhPs while the $k_p$ component is attributed to the edge-launched SPhPs. Following an established procedure, we fit the profiles of the near-field amplitude signal using the equation[65],

$$s_2(\omega) = \left| A \frac{e^{i2k_p x}}{\sqrt{x}} + B \frac{e^{ik_p x}}{x^a} + C \right| \quad (3)$$

where $A$, $B$ and $C$ are complex parameters, $a$ is a real parameter and $k_p$ is the complex-valued wavevector. The first term represents the tip-launched polaritons that travel a distance of $2x$ and the second term corresponds to the edge-launched polaritons that travel a distance of $x$. During the fitting procedure, the real part of $k_p$ is set to the analytically calculated value of the suspended membranes, while the imaginary part is treated as adjustable parameter. The term $\sqrt{x}$ accounts for the geometrical spreading factor of the cylindrical-wave nature of the tip launched polariton while the factor $1/x^a$ represents the efficiency on the excitation of the edge-launched polariton by the synchrotron beam. The fits for the $s_2(\omega)$ profile (Fig.3e) show a good agreement with the experimental data. A high $Q$ factor of up to 23.6 is obtained for the far-infrared SPhPs in the RB$_2$, which closely matches our theoretically calculated $Q$ factor for RB$_2$ (Fig.4d and Fig.1d). Achieving $Q$ factors matching theoretical results underscores the ability of membrane technology to create highly crystalline, atomically smooth, and low-defect membranes, thereby preserving the optical properties of the bulk single-crystalline materials. This high $Q$ factor is comparable to the values of well-known low-loss polaritonic materials, such as the mid-infrared SPhPs in SiC thin films ($Q \sim 28$)[59] and hyperbolic PhPs in hBN ($Q \sim 20$)[4] and MoO$_3$ ($Q \sim 25$)[1,41]. Additionally, the $Q$ factor is significantly larger than that of terahertz hyperbolic PhPs in MoO$_3$ ($Q \sim 7.5$)[37] and GeS ($Q \sim 10$)[35]. This comparison highlights the promise of SrTiO$_3$ membranes as an emerging material platform for advanced far-infrared polaritonic applications. Additionally, we analyze the $k_p$ of mid-infrared SPhPs in RB$_3$, which show significantly shorter propagation length (Supplementary Note 6 and Supplementary Fig.6). These SPhPs demonstrate a low $Q$ of about 5, consistent with our previous study and with the theoretical calculation in Fig.1d.

To demonstrate the importance of suspension, we perform a similar SINS line scan in the region of the 100 nm SrTiO$_3$ membrane supported by the SiO$_2$/Si substrate (Supplementary Note 7 and Supplementary Fig.8). We see that the effect of the substrate is to slightly increase the SPhP confinement factor $C$ (or $k_p$) and to significantly decrease the quality factor, so that the maximum Q factor value in RB$_2$ is only 6. This can be understood by considering how the dielectric function of the substrate and the non-symmetric dielectric environment of the membrane affect the dispersion relations of PhPs. The increase in the confinement factor is primarily attributed to the real part of the substrate's permittivity, while the decrease in the $Q$ factor is attributed to substantial imaginary part of the $SiO_2$[66–68]. Consequently, the implementation of the far-infrared SPhPs necessitates the use of low-loss substrates or suspension of the membrane.

We further demonstrate the thickness tunability of SPhPs in suspended $SrTiO_3$ membranes. Fig.4a shows a frequency-position map of the near-field signal obtained on a 50 nm thick suspended membrane, where multiple SPhP fringes can be seen in $RB_2$. The spacing and spatial extent of these fringes are reduced compared to the 100 nm thick membranes, corresponding to increased confinement (or $k_p$) with $C$ up to 12 (Fig.4c and Supplementary Fig.7). Notably, the 50 nm membrane maintains a high $Q$ factor of up to 22, while the thickness is less than 0.2% of the free-space wavelength (Fig.4d). We also detect the SPhPs in thinner, 30 nm membranes (Fig.4b), although these data show a significantly reduced signal-to-noise ratio due to limited signal from the membrane's reduced thickness. This results in only one fringe in $RB_3$, making the quantitative analysis difficult.

Lastly, we analyze the SPhPs lifetime defined as $\tau = L_p/v_g$, where $L_p = 1/\text{Im}(k_p)$ is the propagation length and $v_g = \text{Re}(d\omega/dk_p)$ is the group velocity. Using $v_g$ obtained from the analytical dispersion (Fig.1c) and experimentally extracted $L_p$ (Supplementary Fig.9), we find $\tau = 8 \pm 1$ ps for the 100 nm membrane and $\tau = 7 \pm 1$ ps for the 50 nm membrane (Fig.4e, Supplementary Note 8). These ultralong lifetimes are comparable to those of the state-of-the-art phonon polaritonic materials, such as mid-infrared PhPs in isotopically enriched hBN ($\tau = 2$ ps)[18], naturally abundant $MoO_3$ ($\tau = 8$ ps)[22,25,69], and freestanding SiC thin films ($\tau = 8 - 9$ ps)[59] as well as terahertz PhPs in $MoO_3$ ($\tau = 9$ ps)[37] and GeS ($\tau = 2 - 3$ ps)[35], further confirming the low-loss nature of SPhPs.

**Conclusion and outlook**
Our study reveals that transferable functional perovskite oxide nanomembranes host low-loss highly confined far-infrared phonon polaritons with the lifetime reaching 8 ps when they are separated from a lossy substrate. Therefore, they represent a promising new platform for long wavelength nano-photonics. The energy of the surface phonon polaritons in the $SrTiO_3$ membrane ranges from 30 to 50 meV, close to room-temperature thermal energy (25 meV), making them easily thermally excitable. Consequently, the discovery of the long-range propagating low-energy surface phonons in the oxide membranes offers significant advantages that higher-energy mid-infrared phonon polaritons in traditional polaritonic materials cannot provide, which will enable a variety of intriguing applications, including sensing, photodetection, enhanced near-field energy transfer, enhanced in-plane thermal conductivity, and polariton-mediated coherent thermal emission. Oxide membranes provide a complementary approach to exfoliable vdW materials, offering distinct advantages such as enhanced chemical and mechanical stability, large lateral dimensions, and potential scalability in production. It is noted that the $SrTiO_3$ used in this work was synthesized from naturally occurring elements, where a notable isotopic variation of strontium and titanium occurs. As demonstrated in the case of hBN[18] and MoO$_3$[70], isotopic enrichment of the membranes may enhance the SPhP propagation length and lifetime even further. Besides SrTiO$_3$, various perovskite oxide membranes can be synthesized, each with unique optical phonon modes and infrared properties. Exploring these membranes could expand capabilities and enable ultra-low loss PhPs, enhancing functionality and performance. Another advantage of the membranes is the ability to control the growth of epitaxial superlattices, where SrTiO$_3$ layers are alternated with other perovskite oxides that have an opposite sign of optical permittivity. The optical property of superlattice membrane can be tailored on demand by selecting functional oxide layers with varying dielectric functions and controlling the thicknesses ratio of the constituent layers. This approach should open avenues to creation of novel optical metamaterials supporting low-loss infrared hyperbolic phonon polaritons, with additional confinement and possibility of engineering the optical density of states in photonic applications.

**Methods**

**Thin-film Synthesis.** Epitaxial heterostructures of 100 nm, 50 nm, and 30 nm SrTiO$_3$ films were synthesized with a 12 nm Sr$_3$Al$_2$O$_6$ sacrificial layer on (001)-oriented SrTiO$_3$ substrates by a RHEED-assisted pulsed laser deposition system (NBM Design). Synthesis of the Sr$_3$Al$_2$O$_6$ sacrificial layer was conducted in vacuum at a pressure of 10$^{-6}$ Torr, temperature of 950 °C, laser fluence of 2.0 J/cm$^2$, spot size of 2.6 mm$^2$, and a laser repetition rate of 1 Hz. Synthesis of the SrTiO$_3$ films was conducted in an oxygen environment of 0.1 Torr, temperature of 760 °C, laser fluence of 1.23 J/cm$^2$, spot size of 2.6 mm$^2$ and a laser repetition rate of 1-3 Hz.

**SrTiO$_3$ Membrane Fabrication.** A 600 nm-800 nm thick polymethyl methacrylate (PMMA) support layer was spin coated on top of the SrTiO$_3$/Sr$_3$Al$_2$O$_6$/SrTiO$_3$ heterostructures and cured at a temperature of 135 °C for 5 minutes. The heterostructures were then placed into deionized water and left at room temperature until the Sr$_3$Al$_2$O$_6$ sacrificial layers were completely dissolved. After releasing the PMMA coated SrTiO$_3$ films from the substrates, the membranes were transferred onto SiO$_2$/Si substrates with arrays of 5 μm wide and 1 μm deep trenches. The trenched substrates were prepared by reactive ion etching (RIE) of photolithographically patterned SiO$_2$. To remove residual water, the transferred membranes were baked at 90 °C for 5 minutes. Finally, the PMMA layers were dissolved in acetone at 60 °C for 30 minutes, rinsed immediately with isopropanol, and dried with nitrogen, leaving only the membranes.

**Materials Characterization**. RHEED patterns and oscillations were recorded at an electron energy of 20 kV using a TorrRHEED system from Staib Instruments and analyzed with kSA 400 software from k-Space Associates. AFM tapping mode images were acquired with an Asylum Research MFP-3D Origin+ AFM using BudgetSensor Tap300DLC tips. X-ray theta-2theta line scans were performed on a Rigaku Smartlab X-ray Diffractometer. SEM images were taken using a Hitachi SU8700 scanning electron microscope. Atomic-resolution STEM images were taken using a probe aberration-corrected ThermoFisher Titan scanning transmission electron microscope operating at 300 keV.

**Synchrotron Infrared Nanospectroscopy (SINS).** SINS experiments were performed at Advanced Light Source (ALS) beamline 2.4. The beamline uses an optical setup consisting of an asymmetric Michelson interferometer mounted into a commercial s-SNOM microscope (neaSNOM, Neaspec GmBH). The setup can be described as an AFM possessing optical access to acquire the optical near-field. The incident IR beam within the interferometer is split into components by a diamond beamsplitter. The split beam defines the two interferometer arms formed by a metallic AFM tip and an IR high-reflectivity mirror mounted onto a translation stage. The IR beam component of the tip arm is focused by a parabolic mirror on the tip-sample region. In the experiment, the AFM is operated in tapping mode, wherein the tip is electronically driven to oscillate in its fundamental mechanical frequency Ω (~250 kHz with an amplitude of ~90 nm) in close proximity to the sample surface. The back-scattered light stemming from the tip-sample interaction is combined at the beamsplitter with the IR reference beam and detected with a high-speed IR detector. A lock-in amplifier using $\Omega$ as a reference frequency demodulates the signal and removes the far-field contributions. The resulting interference signal is Fourier-transformed to give the amplitude signal $s_2(\omega)$ and phase $\varphi_2(\omega)$ spectra of the complex optical signal $S_2(\omega) = s_2(\omega)e^{i\varphi_2(\omega)}$. All SINS spectra were measured with a liquid helium cooled HgCdTe (MCT) detector, equipped with a CsI window, which provides near-field signal down to < 200 cm$^{-1}$. The spectral resolution was set to 4 cm$^{-1}$ for a Fourier processing with a zero-filling factor of 4. All spectra in this work were normalized by a reference spectrum taken from a clean, gold surface.

**Electromagnetic Simulations.** Electromagnetic simulations were performed using a commercial finite-difference and time-domain (FDTD) simulator (Ansys Lumerical). Periodic boundary conditions were used in the *y* direction (see Fig.1e), while PML boundaries were set in the *x* and *z* directions. The source in the simulation was a waveguide port placed on the SrTiO$_3$ membrane. The dielectric function of SrTiO$_3$ used was extracted from our reflectivity measurements.

**Contributions**
Y.L. and R.X. conceived and supervised the study. K.K., R.G. and Y.W. synthesized and characterized the membranes. H.A.B., K.K. and S. G. C. performed the SINS measurements. J.T.-G. and A.B.K. performed the far-field reflectivity measurements to extract the dielectric function of materials. K.K., A.B.K. and Y.L. performed the simulation and data analysis. Y.L., R.X., K.K., A.B.K. and J.T.-G. wrote the manuscript with input from all the authors. All authors contributed to the scientific discussion.

**Data availability**
All data that support the findings of this study are present in the paper and the Supplementary Information. Further information can be obtained from the corresponding author upon request.

**Acknowledgements**
The authors thank Xiaobin Xu for valuable discussions. Y.L. and Y.W. acknowledge the support by the National Science Foundation under Award No. DMR-2340751. R.G. and R.X. acknowledge the support by the National Science Foundation under Award No. DMR-2442399 and the American Chemical Society Petroleum Research Fund under Award No. 68244-DNI10. This research used resources of the Advanced Light Source, a U.S. DOE Office of Science User Facility under contract no. DE-AC02-05CH11231. The research of J.T.-G. and A.B.K. is supported by the Swiss National Science Foundation (grants TMPFP2_224378 and #200020_201096).

# Figures

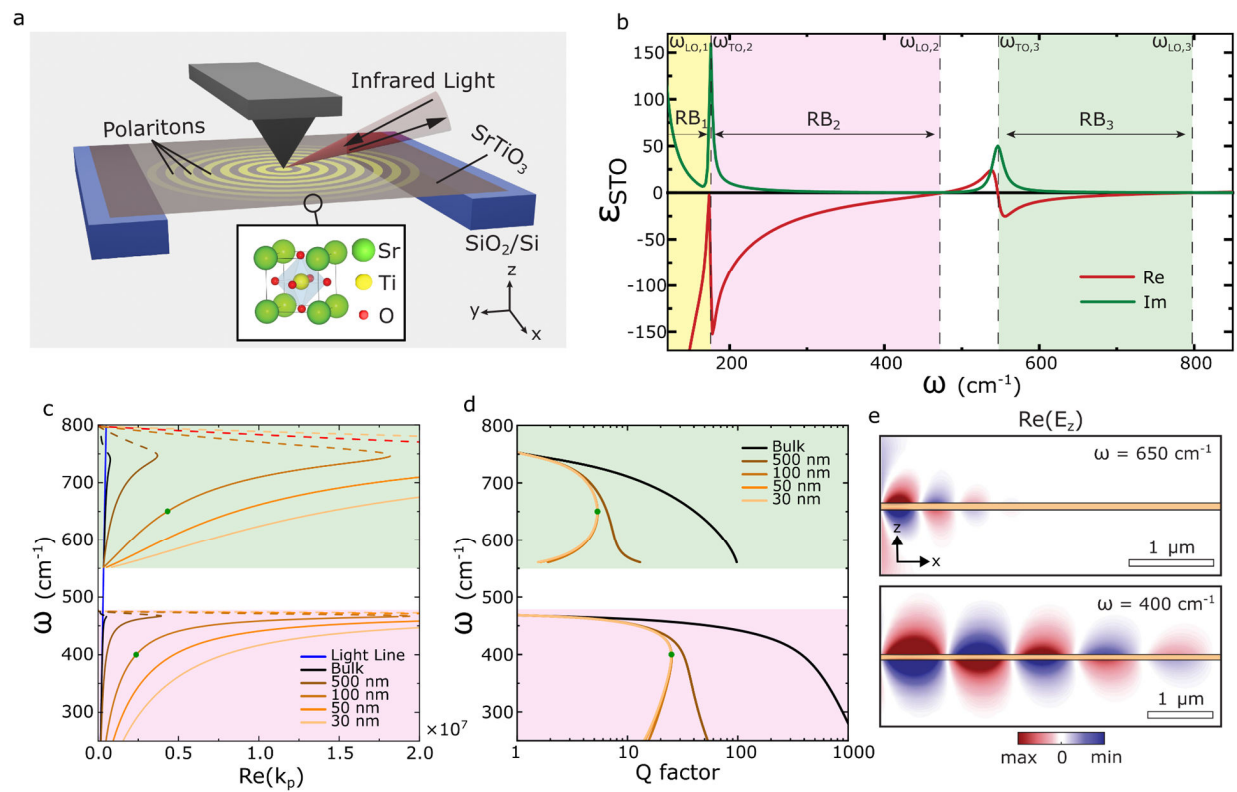

**Figure 1. Low-loss far-field surface phonon polaritons (SPhPs) in suspended SrTiO$_3$ membranes**. **a)** Schematic of the s-SNOM/SINS measurement of suspended SrTiO$_3$ membrane. **b)** Experimentally extracted optical dielectric function of SrTiO$_3$ with labeled Reststrahlen bands (RBs). **(c,d)** Calculated dispersions (c) and figure of merit $Q$ (d) of SPhPs in suspended SrTiO$_3$ membranes with different thickness. RB$_2$ and RB$_3$ are marked in pink and green respectively. Symmetric and antisymmetric SPhPs are denoted by dashed and solid curves in (c) respectively. The green dots in **c,d** correspond to the two frequencies used for the field distribution simulation in (e). **e)** Simulated electric field distributions of SPhPs at 650 cm$^{-1}$ (in RB$_3$) and 400 cm$^{-1}$ (RB$_2$) respectively.

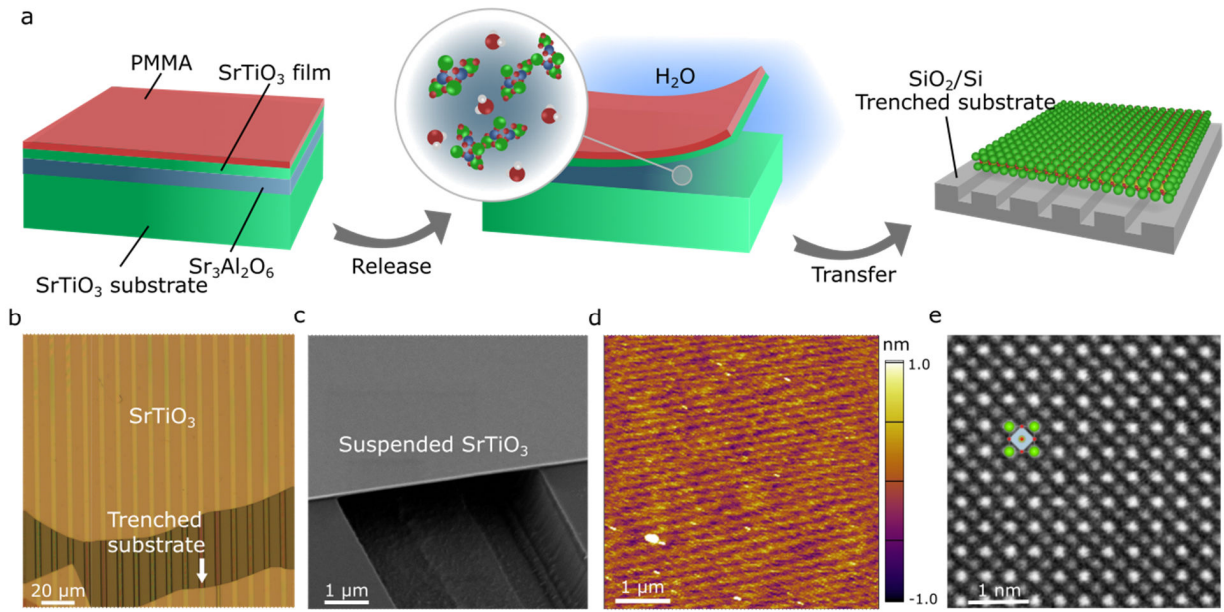

**Figure 2. Fabrication and characterization of SrTiO₃ membranes**. **a)** Schematic illustrating the synthesis and transfer of SrTiO₃ membranes onto trenched substrates for the suspension. **b)** Optical microscope image of a membrane with lateral size of about 5 millimeters transferred on the trenched substrate. The arrow marks a sharp edge perpendicular to the trench. **c)** SEM image showing the suspended membrane with a sharp edge, clean surface. **d)** AFM image of transferred membrane with step terraces, indicating atomically smooth surface. **e)** Atomic scanning transmission electron microscopy (STEM) image of SrTiO₃ membrane with an overlay of the unit cell of a perovskite structure.

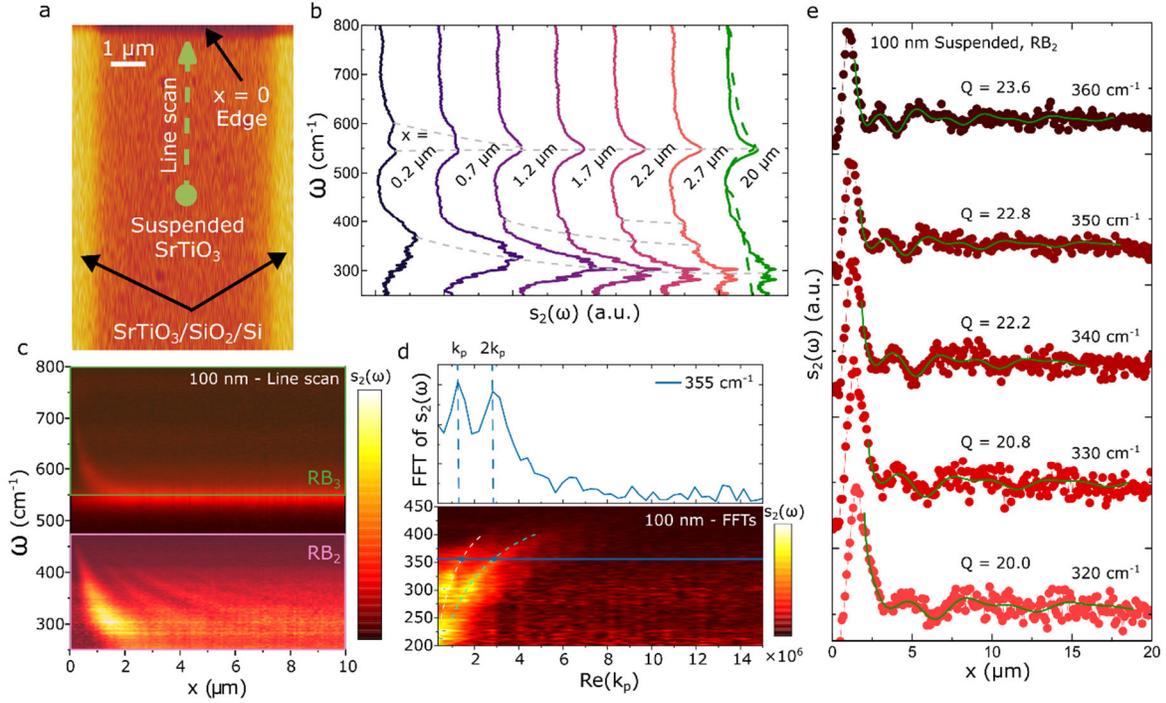

**Figure 3. SINS nanospectroscopic analysis of a 100 nm thick suspended SrTiO₃ membrane.**
**a)** Near-field optical image of the suspended (orange) and supported (yellow) membrane. The green dashed line indicates the SINS line scan direction. The sharp, perpendicular, membrane edge is used as the x=0 μm reference in plots (a-e). **b)** SINS spectra acquired at various distances from the sample edge. Green (solid) spectra is a SINS spectrum taken far away from sample edge, green (dashed) spectra is a SINS spectrum simulated by finite dipole modeling. **c)** SINS amplitude spectra ($s_2(\omega)$) obtained by a line scan perpendicular to the edge of the membrane, showing SPhP fringes in RB$_2$ and RB$_3$. **d)** Fourier transform spectra of a SINS spectrum at 355 cm$^{-1}$ (top) and at all frequencies in RB$_2$ (bottom), showing the line profiles are dominated by two spatial frequencies corresponding to k$_p$ and 2k$_p$. In the bottom panel, white dashed lines correspond to analytically calculated dispersion of 100 nm in Fig.1d and blue horizontal line marks the line cut at ω = 355cm$^{-1}$, corresponding to the spectrum in the top panel. **e)** SINS amplitude ($s_2(\omega)$) line profiles (symbol) at different frequencies in RB$_2$. The green curves are the fit of the experimental data with the equation, $S(x) = A\frac{e^{i2k_px}}{\sqrt{x}} + B\frac{e^{ik_px}}{x^a} + C$ for the extraction of the figure of merit $Q$.

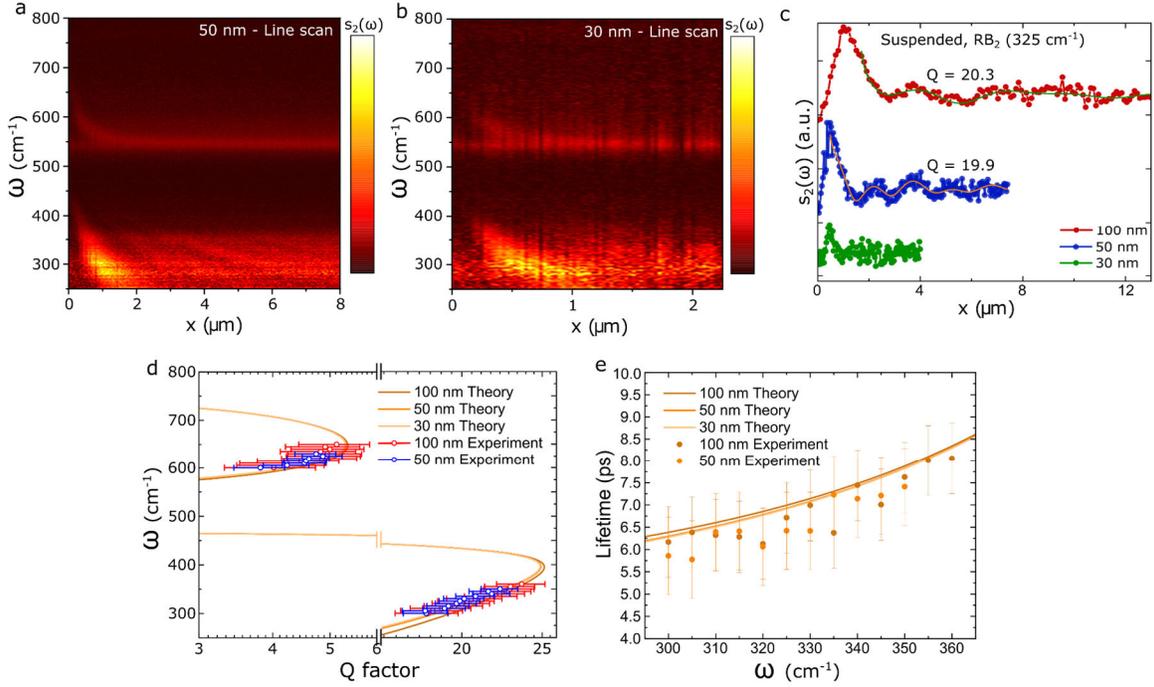

**Figure 4. Thickness-dependent tuning of SPhPs in suspended SrTiO$_3$ membranes**. (**a,b**) SINS amplitude spectra were obtained through line scans on membranes with thicknesses of 50 nm (a) and 30 nm (b), respectively. The line scans are performed perpendicular to the edge of the membranes (at x=0). **c)** SINS amplitude line profiles (symbol) at $\omega$=325 cm$^{-1}$ in RB$_2$ for membranes with different thickness. The green and red curves are the fit of the experimental data with the equation for the extraction of the figure of merit $Q$. **d)** Experimentally extracted (symbols) and theoretically calculated (solid curves) $Q$ factors of the SPhPs in membranes with different thickness. The error bars represent the 95% confidence interval. **e)** lifetime $\tau$ of SPhPs in suspended SrTiO$_3$ with varying thicknesses. Solid curves represent calculations, while symbols indicate experimentally extracted values.